\documentclass[12pt, technote, onecolumn]{IEEEtran}
\usepackage{cite}

\ifCLASSINFOpdf
\usepackage[pdftex]{graphicx}
\graphicspath{{./}}
\else
\fi
\usepackage{url}

\usepackage{multirow, array}
\usepackage{rotating, tabularx}
\usepackage{longtable}
\usepackage{placeins}

\usepackage{xspace}
\usepackage{eurosym}

\newcommand{\quong}{QUonG\xspace}

\newcommand{\apenetp}{APEnet+\xspace}

\newcommand{\ie}{\textit{i.e.}\xspace}
\newcommand{\eg}{\textit{e.g.}\xspace}
\newcommand{\etc}{\textit{etc.}\xspace}

\newcommand{\euretile}{EURETILE\xspace}
\newcommand{\lofamo}{LO{\textbar}FA{\textbar}MO\xspace}

\begin{document}
%
\title{'Mutual
  Watch-dog Networking': Distributed  Awareness of Faults 
  and Critical Events in Petascale/Exascale systems}



%
\author{\IEEEauthorblockN{Roberto Ammendola\IEEEauthorrefmark{1}, 
Andrea Biagioni\IEEEauthorrefmark{2},
Ottorino Frezza\IEEEauthorrefmark{2}, 
Francesca Lo Cicero\IEEEauthorrefmark{2},}
\IEEEauthorblockN{Alessandro Lonardo\IEEEauthorrefmark{2},
Pier Stanislao Paolucci\IEEEauthorrefmark{2},
Davide Rossetti\IEEEauthorrefmark{2},
Francesco Simula\IEEEauthorrefmark{2},  
Laura Tosoratto\IEEEauthorrefmark{2} and  
Piero Vicini\IEEEauthorrefmark{2}}\\

\IEEEauthorblockA{\IEEEauthorrefmark{1}INFN Sezione di Tor Vergata,
Rome, Italy\\ Email: [firstname.lastname]@roma2.infn.it}\\

\IEEEauthorblockA{\IEEEauthorrefmark{2}INFN Sezione di Roma,
Rome, Italy\\ Email: [firstname.lastname]@roma1.infn.it}}
\maketitle



%
\IEEEpeerreviewmaketitle

\section{Introduction}
Features like resilience, power consumption, and availability of large
scale computing system strongly depend on \emph{1-} the complexity of
individual components (\eg the gate count of each chip) and \emph{2-}
the number of components in the system.
Exa-scale computing systems and networks of 3G devices are examples of
distributed systems composed of a huge number of high complexity
individual devices.
Indeed, the FIT (Failures In Time) rate of individual hardware
components, when scaling to peta- and exa-scale systems, becomes a
hard challenge to the collective efficiency or even reasonable
functionality for the platform.

A first step to mantain systemic functionality and efficiency is the
adoption of hardware design techniques which improve the individual
component.
Considering the case of fault resilience, multi-core sockets used on
many-core systems must adopt resiliency techniques to reduce the FIT
(Failures In Time) rate. This design trend is already clear in the
transition between past tera-scale systems, adopting commodity
processors with $0.1\div0.5$ fails per year per socket, to peta-scale
systems, where the failure rate could be reduced to 0.001 fails per
year per socket\cite{Kogge} adopting hardware design techniques like
memory and bus encoding, memory scrubbing, provision of spare
processors and memories. Even considering those reduced FIT rates and
a very limited number of components, mission critical/life support
systems mandate for architectures adopting double or triple
redundancy.
In practice, petaFLOPS designs based on resilient sockets adopting
such countermeasures are characterized by a rate of system-stopping
features in the range of a few days, while a system failure rate in
the range of few hours is displayed by systems mounting less resilient
sockets.\cite{Elnozahy}.

Without additional measures, the FIT rate of exa-scale systems becomes
unacceptable due to the scaling in the number of components.
Analogously, for what regards power and thermal issues, each socket
and component is nowadays designed keeping the energetic concerns as
key drivers but systemic countermeasures are required due to the
numerosity of components.

In our vision, a necessary feature on larger scale architectures is
the detection and collation of relevant information about faults and
critical events and, due to the distributed nature of the system, the
reliable propagation of this awareness up through the system
hierarchy. In other words, the system must be rendered
\textit{fault-aware} to be able to choose and enact the actual system
fault response.

Based on these considerations, the \euretile project starts bottom-up
proposing a mechanism that creates a systemic awareness of fault and
critical events, the \lofamo design: a distributed, mutual watch-dog
paradigm which enables the management of faults and critical events in
large scale systems.

The \lofamo design can complement a pro-active mitigation approach,
\ie the enforcing of preventive actions (\eg preemptive process
migration, dynamic clock frequency reduction for individual
components, dynamic routing strategies, \etc) before the failure
occurs, so as to avoid faults that can reasonably be expected or
minimize the impact of those who can not.

In this paper, we will mainly focus on the requirements imposed by \emph{1-}exa-scale systems\cite{exa}, \ie an assembly of tens or hundreds of
thousands of processors, hundreds of I/O nodes and thousands of disks,
and \emph{2-} future many-tile sockets; however, similar techniques
could be applied to large networks of independent, autonomous
devices.  

Our approach should mitigate the performance penalty and productivity
reduction due to work loss and failure recovery, obtainable using
exclusively the conventional approach to fault tolerance
(checkpointing/failure/rollback), which is foreseen to be
problematic\cite{Kogge}.

With the assistance of some hardware components located in the DNP
(Distributing Network Processor), the router of our \apenetp, the
\lofamo design paradigm employs some 'watchdog' techniques for
reciprocal fault diagnosis between the DNP itself and a companion
'host processing element' within a node or on nodes which are
neighbouring in the \apenetp mesh topology; moreover, it employs a
number of \textit{best-effort} heuristics for delivering the
diagnostic data even in case of faulty or broken links along an
auxiliary support network. This network is a possibly low-speed, but
highly reliable, diagnostic-dedicated, independent one, which leans on
its high-speed 3D toroidal companion mesh in the extreme case of its
own failing.

Once complemented with diagnostic facilities that monitor whatever
metrics are deemed relevant to the prediction of faults (\eg
temperature and voltage/current probes, BER counters, \etc), \lofamo is a
keystone upon which a fault management system can draw inferences that
drive its strategies and actions to keep the system up and running.

Future work is aimed towards increasing the capabilities of the DNP
regarding local re-routing, in sight of the most ambitious goal of
building a network mesh whose routing is deadlock-free and
fault-tolerant.

\section{Definitions}
In this section we establish a basic terminology to describe a fault
(and critical event) tolerant system, which will be used in the
following sections of this document. First of all, from now on we will
use \emph{fault} as abbreviation for \emph{fault and critical
  event}. Then, we split the fault-tolerance problem in two major key
areas: \emph{fault awareness} and \emph{fault reactivity}:
\begin{itemize}

\item{\textbf{Fault awareness}} is the capability of the system to
  assess its own health status, in order to acknowledge faults that
  have already appeared or to make guesses about those likely to
  occur. Going bottom-up, this 'introspection' can be reduced to two
  aspects:

\begin{itemize}
\item{\textbf{Local fault detection},} the capability of a device to
  perform a number of HW and SW tests to detect a condition of fault
  in itself or other contiguous devices.
\item{\textbf{Systemic fault awareness},} the collation of diagnostics
  propagated throughout the whole network by the local detecting
  sub-systems to compose a global picture of the system's health.
\end{itemize}

\item{\textbf{Fault reactivity}} is the range of initiatives that the
  system enacts, under the presumptions it can make when its own
  global health is known to it, to prevent a fault situation which is
  about to occur or to gracefully degrade its performance instead of
  bringing the whole system to a stop when the fault has
  occurred. Going top-down, this 'self-adjustment' can be reduced to
  two aspects:

\begin{itemize}
\item{\textbf{Systemic response},} the set of strategies that the
  system can choose to apply, following inferences that it can make
  from its own diagnostic self-image, to prevent and counter the
  faults.
\item{\textbf{Local readjustment},} the set of readjustments that can
  be locally enacted to prevent and counter the faults, \eg reduction
  in clock frequency, changes to the routing tables to bypass a faulty
  link, remapping the assignment of tasks to nodes, \etc.
\end{itemize}

\end{itemize}
\begin{figure}[!hbtp]
\centering
\includegraphics[width=.8\textwidth]{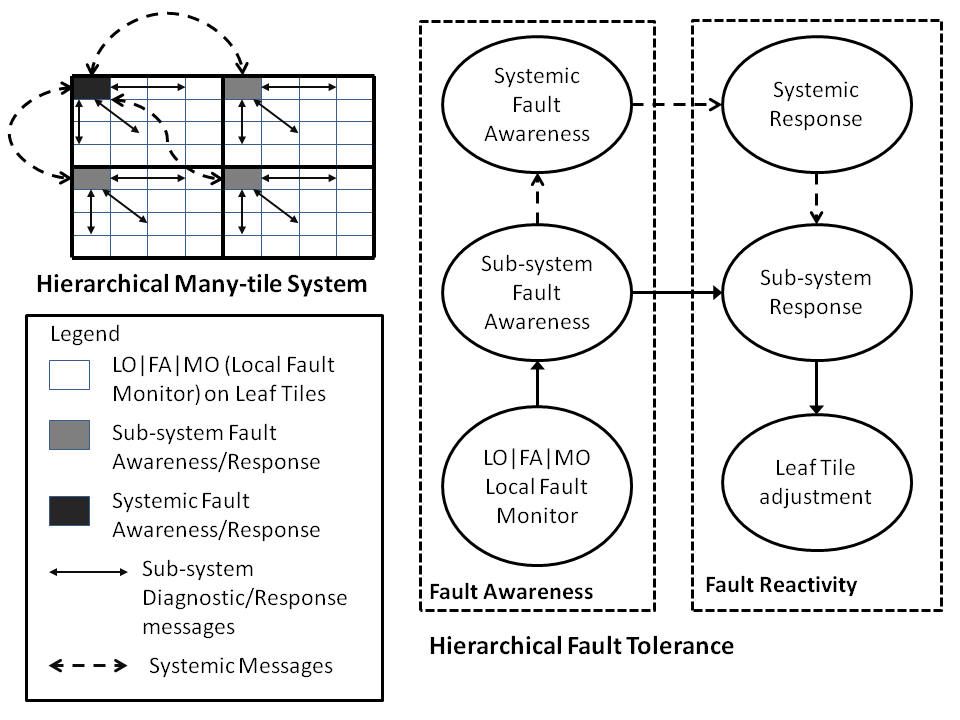}
\caption{The Network Processor of each leaf in the many-tile HW system
  is equipped with its own \lofamo components. The local Fault
  Awareness is propagated towards the upper hierarchy levels, creating
  systemic Fault Awareness. Fault reactions could be autonomously
  initiated by sub-system controllers or could require a systemic
  response.}
\label{fig:LOFAMO_hierarchy}
\end{figure}

It is clear that a complete design of a fault-tolerant architecture
must give detailed specifications in each of the abovementioned areas.
On the other hand, the challenging part for the most interesting
fault-tolerant features is the actual implementation, which cannot be
detached from a low-level specification of the host architecture.
For example, task migration capabilities are derived from process
management features of the host operating system; application
checkpointing is strictly bound to storage options available to the
host node; protection from memory errors by ECC is a low-level
addition to the host memory architecture, \etc.

Our idea with \lofamo is that of a fault-tolerant framework which is
as host-agnostic as possible. By encapsulating as many features as can
be accommodated independently from such host specifications in
\lofamo, we strive to achieve a clear separation of problems - with
the hope this leads to easier solution - and a degree of design reuse.

By saying this, we make clear from the start that \lofamo, by its very
nature, has to be restricted to the side of fault-awareness.

\section{Fault awareness: global scenario hypotheses}
We give here a sketch of an architecture where \lofamo is employed. We
assume a computing mesh where every node is a combination of the
hardware supporting \lofamo, \ie a DNP (Distributing Network
Processor), mated to a host processing element; beyond the
communication facilities provided by the DNP, the host exposes another
communication interface towards an auxiliary 'service network' (more
details in section \ref{sec:ServiceNetwork}).

Failures can generally be of \textit{commission} and \textit{omission}
type: the former encompasses the case of failing elements performing
their tasks in an incorrect or inconsistent way (\eg corruption in
node memory, corruption in transmitted messages, packet misrouting,
\etc); the latter deals with the case of failing elements skipping
their tasks altogether (\eg node stops responding due to crash
failure, power outage or burn-out, message passing does not progress
due to link disconnection, \etc).

The most general kind of faults are those where the behaviour of a
faulty component is assumed to possibly be completely random as to its
correctness; in literature, fault-tolerance to this kind of faults is
defined \textit{Byzantine} fault-tolerance\cite{Byzantine}.
\textit{Byzantine} failures can be seen as undetectable commission
failures or, where possible, as malicious activity by some agent which
is trying to sabotage the network. This kind of failures is explicitly
not covered here.

With this restriction, detectable commission failures signal either a
component that is about to break or keeps on working wrong, while
omission failures, when permanent, mostly stand for an already broken
or disconnected component.

In this picture, the \lofamo design is charged of polling the supplied
sources of diagnostic data; any inconsistent value, be it any value
beyond a certain threshold or a timed-out update of a watchdog
counter, is a failure to report.
\lofamo attempts then to push this report along the service network -
which means, in emergency cases, leaning against the neighbouring
DNP's - towards an upper layer Fault Supervisor.

As per previous definitions, we remark that the only behaviours
\lofamo foresees for a failing component are two:
\begin{itemize}
\item{\textit{sick} - the component has a rate of detected commission
  failures beyond the compatibility threshold of normal operativity
  $\rightarrow$ this \emph{may} need action};
\item{\textit{failed} - the component has a permanent commission
  failure (it keeps on working wrong) or simply stops participating in
  the network, \ie it has a permanent omission failure (it has broken)
  $\rightarrow$ this \emph{needs} action}.
\end{itemize}

\subsection{EURETILE Platforms terms and definitions}
Here we introduce a few concepts about the EURETILE platforms and we
define the terms used in the following sections to refer to the
platform components.\\

\begin{figure}[!ht]
\centering
\includegraphics[width=0.8\textwidth]{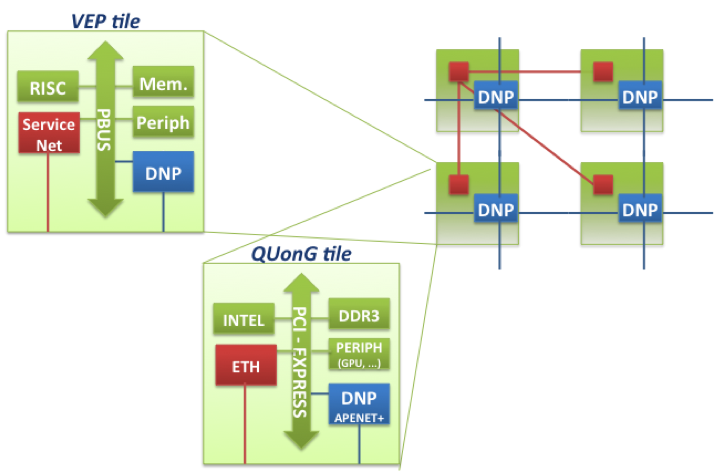}
\caption{The EURETILE platforms.}
\label{fig:VEP_QUONG}
\end{figure}

\noindent
\begin{longtable}{p{0.2\textwidth}p{0.75\textwidth} }
EURETILE\newline architecture & A many-tile system, where the
elementary HW tile is a multi-processor, which includes a Distributed
Network Processor (for inter-tile communication), a Host processor
(for control, user interface and sequential computations), and a
floating-point numerical engine (for high intensity arithmetic
computations), potentially disjointed from the Host processor.\\ \\
DNP & Distributed Network Processor, the component implementing the 3D
torus interconnection network between the tiles, providing RDMA
support for data transfers. The two implementations of the DNP are:
\begin{itemize}
\item \textbf{\apenetp}, an FPGA-based card for low latency, high
  bandwidth direct network interconnection, supporting
  state-of-the-art wire speeds and providing a PCIe X8 Gen2
  interface\cite{ApenetLattice10};
\item \textbf{DNP-VEP}, a SystemC TLM model of the DNP for the Virtual
  EURETILE Platform (see below).
\end{itemize}
The \emph{DNP Core} is the internal DNP logic, including the routing
logic and the RDMA engine; it does not include the links and the Bus
interface. The functionalities of the DNP core are likewise
implemented both in DNP-VEP and in \apenetp.\\ \\
Host & The processor in the tile running the OS and interfacing with
the DNP and the peripherals through a Bus.\\ \\
Virtual EURETILE Platform & The EURETILE simulation platform,
integrated with the SW toolchain and available to run applications and
to collect profiling data. This platform implements the EURETILE
architecture with a basic tile that includes: a RISC-like Host
processor, a SystemC model of the DNP (DNP-VEP), an external memory
and a number of peripherals.\\ \\
\quong HPC Platform & The EURETILE demonstration hardware platform,
whose basic tile includes: an x86-64 multiprocessor as Host, an
\apenetp card and a GPU as floating-point accelerator.
\cite{QuongSaahpc}.\\
\end{longtable}
In figure \ref{fig:VEP_QUONG} the two flavours of the EURETILE tile
are schematically shown; we stress on the fact that their connectivity
is ensured either by the DNP to create the 3D torus topology and by a
Service Network for the diagnostic purposes described in section
\ref{sec:ServiceNetwork}.
\FloatBarrier
\subsection{Local fault monitor} 
\begin{figure}[!htbp]
\centering
\includegraphics[width=0.9\textwidth]{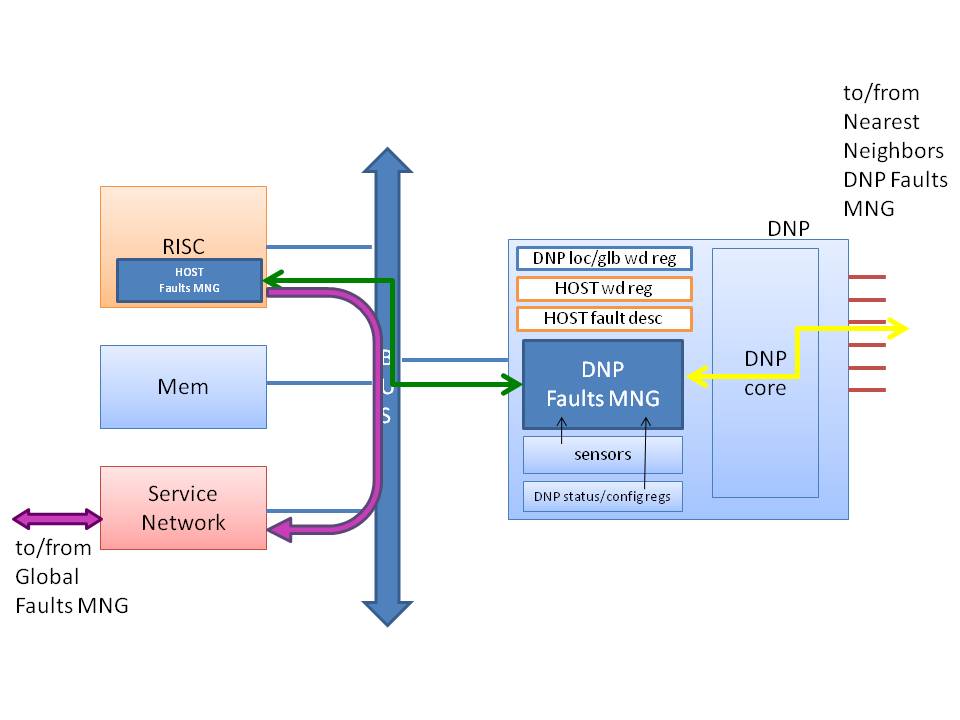}
\caption{The basic EURETILE platform configuration showing the
  position and the communication paths of the \emph{Host Fault
    Manager} and the \emph{DNP Fault Manager} in a \lofamo
  implementation. The Host watchdog register stores info about Host
  Faults. The DNP watchdog register stores info about local or global
  faults of the networking system. The Host Remote Fault Descriptor contains information about the nature of remote Host Faults.}
\label{fig:LOFAMO}
\end{figure}

The Local Fault Monitor (\lofamo) is the mechanism chosen to obtain
the fault-awareness; it implements health self-tests for a number of
hardware devices and takes care of propagating the deriving
information. Moreover, the devices are able to monitor other
contiguous devices and communicate their faulty status.\\
Synthetically, each device is able to:
\begin{itemize} 
\item check/elaborate/store/transmit its own status;
\item monitor other contiguous devices.
\end{itemize}
In the \euretile platform, the actors of the described mechanism are
the DNP/\apenetp and the Host sub-system (Intel for the \quong
platform, a RISC-like model for the Virtual \euretile Platform).
The DNP is able to run self-tests on its own links and logic, as well
as to retrieve information from its own  temperature and
electrical sensors. All information pertaining to the sub-systems status is
gathered by the \lofamo-appointed component inside the DNP itself and
stored in a DNP WatchDog register (see \ref{StatRegs}). A second
register inside the DNP is dedicated to the surveillance of the health
status of the Host, with \lofamo performing periodic checks of the
Host WD register.\\
In the event the Host on one or more nearest neighbouring nodes were
faulty, a third register, the Host Remote Fault descriptor register,
would end up containing information about the nature of the remote
fault.
The self-test capabilities of the DNP links and logic allow mutual
monitoring between nearest neighbour DNP's, all of them acting as
watchdog for one another.\\
The key points for this \lofamo implementation are:
\begin{itemize}
\item the presence of a \textbf{Host Fault Manager} (Host FM), a
  software process running on the Host that is aware of the Host local
  status, is able to read/write the DNP internal registers and the
  \emph{DNP local/global} and \emph{Host watchdog registers} and can
  send messages through the Service Network.
\item the presence of a \textbf{DNP Fault Manager} (DNP FM), a
  component residing on the DNP that is able to collect the information about the DNP health
  status, to read/write the \emph{DNP local/global} and \emph{Host
    watchdog registers} and to send messages through the 3D Network.
\end{itemize}
Figure \ref{fig:LOFAMO} illustrates the basic platform configuration
detailing the position and the communication paths of the Host Fault
Manager and the DNP Fault Manager.\\

Keeping to the definitions given above, the task of \lofamo thus
encloses the whole of Local fault detection and the interface to the
Fault Awareness system.
\subsection{Fault Supervisor}
The Fault Supervisor is the generic term that encompasses the set of
processes receiving the output of the \lofamo machinery; its duty is
to create systemic Fault Awareness and to issue appropriate systemic
Fault Responses.\\
For a small number of nodes, the Fault Supervisor could be implemented
as a single software process running on an appointed 'master node' of
the system; for larger systems, a process cloud residing on a subset
of nodes participating in a hierarchy would certainly be more
scalable.\\
The Fault Supervisor is kept timely fed by the set of DNP Local
Fault Managers and Host Local Fault Managers, with periodic updates about their health.\\
This supervisor is the 'systemic intelligence' that embodies the fault
awareness for the system and drives its response; all information (or
lack thereof, in case of omission failures from faulty nodes, which is
information as well) is brought by the \lofamo network to the Fault
Supervisor system, so that it can choose any fault prediction,
prevention and reaction strategy it deems feasible.\\
The Fault supervisor is a critical component about which, in the
following, we dismiss to provide any more details. This document describes only the \lofamo mechanism and its specific implementation on the QUonG and VEP Platforms; we acknowledge
its presence - after all, it is the target of all communications from
\lofamo - but we are agnostic about anything regarding its
implementation.

\subsection{Service network}
\label{sec:ServiceNetwork}
Besides \apenetp's high speed 3D mesh, \lofamo expects the system
nodes to partake in a secondary, diagnostic-dedicated network fabric
to which only the Host has access. In ordinary conditions, the DNP
relays the gathered diagnostic data to its Host companion which,
through this network, in turn relays them to the Fault Supervisor.
In this way, the high speed network is unencumbered from dealing with
the added traffic of the health status reports.

We expect this service network to be a relatively inexpensive local
interconnect. On the HPC market, Ethernet is
a mature technology, mostly ubiquitous presence for any architecture
we think to match the \apenetp board with - \eg, our \quong platform
cluster node prototype is a
Supermicro\textsuperscript{{\textregistered}} board equipped with dual
Gbit Ethernet. For this reason, the presence of such service network
is regarded as a rather unconstraining addition on the HPC flavour of the EURETILE architecture. On many-tile embedded systems, represented in our case by the VEP platform, we maintain at this stage open the definition of the service network. 

We are positing that the bulk of diagnostic data does need neither
high bandwidth nor extremely low latency. This means that performance
concerns are not overtly constraining in the building of this service
network and all effort can be instead put in pushing its reliability,
by means of ruggedness of components (for the switches, routers, NICs,
cabling, \etc) or some kind of redundancy; reliable Ethernet is a wide
ranging subject with many possible approaches\cite{ReliableEth}.

However, this diagnostic network is a system element itself subject to
failure. So, the problem must be raised of how to deliver fault
awareness data in presence of criticality of the service network
itself or the DNP's.
First, we analyze the case of Host and DNP not affected by
simultaneous fail, then the case of simultaneous fail of the DNP and
Host on a tile.  The hypothesis we put forward is that the probability
for a node of Host and DNP simultaneously failing is significantly
smaller than their individual failure rate. This means that having the
host and the DNP mutually cross-checking each other, \lofamo has
meaningful escape routes in both of the following scenarios:
\begin{itemize}
\item the DNP breaks down $\rightarrow$ the DNP does not respond to
  queries from the Host - the Host acknowledges the omission fault and
  signals it via the service network to the Fault Supervisor (this
  does not differ from the ordinary condition).
\item the Host breaks down $\rightarrow$ the Host does not respond to
  queries from the DNP - although from that node the service network
  is inaccessible, the DNP has a last chance of relaying its reports
  along to its neighbours in the high-speed 3D mesh and from there,
  all receiving DNP's can relay the data to their own Host and then on
  to the Fault Supervisor.
\end{itemize}

Moreover, even in the showstopping event of both Host and DNP breaking
down in a node, the system has a way to become aware of the situation:
no more activity from the node means that all the neighbouring nodes
in the 3D mesh become eventually aware of a permanent omission fault
in one of their channels; as soon as reports of this fact reach the
Fault Supervisor, this latter can infer the node has died and take
relevant action.

\subsection{Watchdog implementation}
\label{Watchdog}
One of the foundation of the \lofamo design is the \emph{mutual
  watchdog mechanism}, that for the EURETILE platform is implemented
in the following way: the DNP acts as watchdog for the Host, \ie it
periodically monitors the Host status as reported in the Host watchdog
register updated by the Host itself; the Host acts as watchdog for the
DNP, \ie it periodically monitors the DNP status as reported in the
DNP Local/Global watchdog register updated by the DNP itself.
Although both the mentioned registers are located inside the DNP they
are written (updated) and validated by their 'owner' and read and
invalidated by the other device.  Validation/invalidation consists of
setting the Valid Bit to 1 or 0, respectively.  The update period is
such that $T_{write} < T_{read}$, in this way is guaranteed that the
reader always founds a valid status and viceversa, unless a
destructive omission fault occurs that makes the writer unable to
update its status register (see section \ref{FaultDetSignal}).
\begin{figure}[!ht]
\centering
\includegraphics[width=0.8\textwidth]{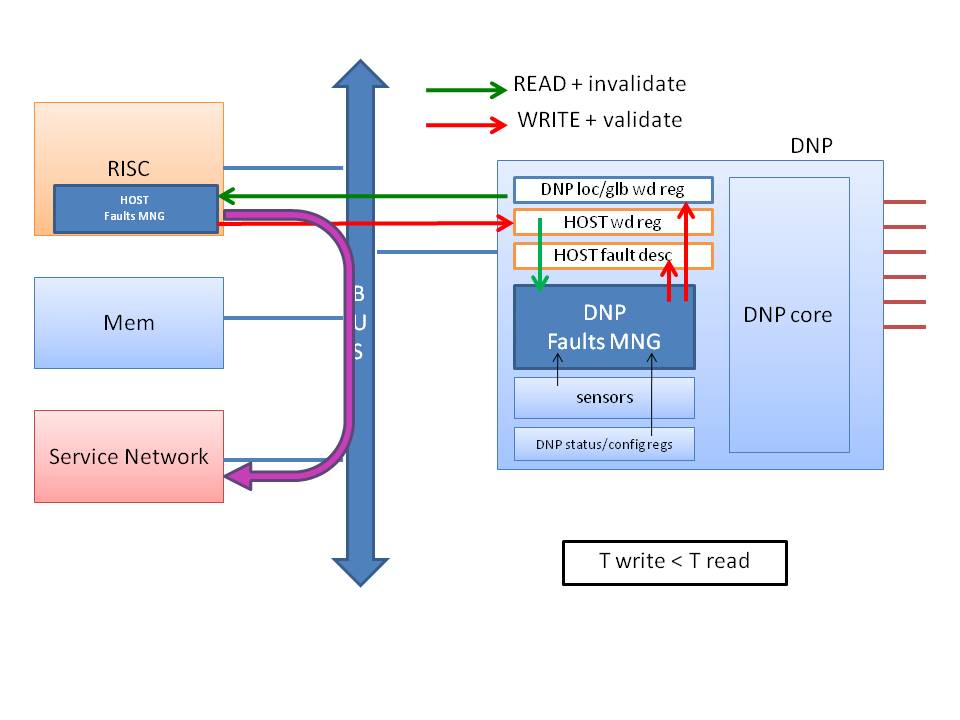}
\caption{The Host and the DNP act as reciprocal watchdogs. 
Even if both the Host and DNP WD Registers are located 
inside the DNP, they are periodically updated by their 
'owner' and invalidated by the other device.}
\label{fig:LOFAMO_wd}
\end{figure}
\FloatBarrier
\begin{table}[!h]
\centering
\setlength\extrarowheight{10pt}
\begin{tabular}{|m{0.7cm}|m{0.9cm}|m{2cm}|m{4cm}|m{4cm}|m{4cm}|}
\hline
\hline
\textbf{Fault class} & \textbf{Faulty Elem.} & \textbf{Faults} & \textbf{Detector} & \textbf{Diagnostic info storage} & \textbf{Diagnostic info path}\\
\hline
\hline
\multirow{7}{0.8cm}{DNP side} & \multirow{2}{0.8cm}{Link} & Link sick
(CRC errs) & receiving Link self-test $\rightarrow$ receiving DNPfm & \multirow{2}{4cm}{bits in
DNP loc/glb wd register} & \multirow{5}{4cm}{DNPfm $\rightarrow$ Host $\rightarrow$
Service Net}\\
\cline{3-4}
& & Link broken &  DNPfm on both the receiving and transmitting side (if link logic still working), otherwise on a single side & & \\
\cline{2-5}
& \multirow{3}{0.8cm}{DNP $^{\circ}$C/W/V alert} & Temperature over/under threshold & \multirow{3}{4cm}{Sensors
$\rightarrow$  DNPfm} & \multirow{3}{4cm}{bits in DNP
  loc/glb wd register}& \\
\cline{3-3}
&&Power over/under threshold &&&\\
\cline{3-3}
&&Voltage over/under threshold &&&\\
\cline{2-6}
& \multirow{2}{0.9cm}{DNP logic} & DNP core sick & DNPfm & bits in DNP loc/glb wd reg
&\multirow{3}{4cm}{Host $\rightarrow$ Service Net}\\
\cline{3-5}
& & DNP core meltdown & Host (DNPfm misses to update DNP
loc/glb wd register) & bit in DNP loc/glb wd reg
&\\
\cline{1-5}
\multirow{3}{0.8cm}{Host side}&\multirow{3}{0.8cm}{Host
  fault}&\multirow{2}{2cm}{Memory, Peripherals (\ldots, Service Net)}&
\multirow{2}{4cm}{Machine dependent but managed by the Host\_fm}&\multirow{2}{4cm}{Spare room in HOST
  watchdog register}&\\
\cline{6-6}
&&&&&\multirow{2}{4cm}{DNP $\rightarrow$ 3D net $\rightarrow$ Neighbour
  DNPfm $\rightarrow$ Neighbour Host $\rightarrow$ Service Net}\\
\cline{3-5}
&&Bus or total Host breakdown & DNPfm (Host misses to update HOST wd
reg)& bit in RISC wd reg &\\
\hline
\hline
\end{tabular}
\caption{A summary of how faults are detected (which component is
responsible to detect them) and how the information is conveyed
bottom-up to obtain the Global Fault
Awareness. (Abbreviations: DNPfm = DNP fault manager; Host\_fm= Host
fault manager; Net = Network; loc/glb = local/global; wd = watchdog)}
\label{tab:FaultDetect}
\end{table}
\FloatBarrier
\clearpage
\section{Implementation Details}

\subsection{\lofamo Status Registers}
\label{StatRegs}
There are three key registers used by the \lofamo components:
\begin{itemize}
\item \textbf{DNP Local/Global watchdog register}. It contains information about: 1- the status of the local DNP and 2- the status of the Hosts on
first neighbouring tiles;
\item \textbf{Host watchdog register}. It contains the local Host status;
\item \textbf{Host Remote Fault descriptor register}. In case of one or more Host(s) on  first neighbour tiles are faulty, it contains information about the nature of the fault. 
\end{itemize}
For the complete description and layout of these registers refer to the tables \ref{tab:DNPwdreg}, \ref{tab:HOSTwdreg},
\ref{tab:HOSTremfault}, while in section \ref{FaultDetSignal} their
use is detailed.
\\
Table \ref{tab:TempValues} provides the encoding for the
\apenetp Temperature register and the corresponding value for the FPGA temperature as measured by the internal sensor.
\FloatBarrier
\begin{table}[!ht]
\centering
\setlength\extrarowheight{3pt}
\begin{tabular}{|c|c|c|l|}
\hline
\hline
\multicolumn{4}{|c|}{\textbf{DNP local/global watchdog register}}\\
\hline 
\hline
\textbf{\# bits} & \textbf{bit range} & \textbf{field name} & \textbf{Description}\\
\hline
1 & 0 & Valid & 1 : register content valid; 0 : register content not valid\\
\hline
1 & 1 & Host Z- & \multirow{6}{*}{Status of the first neighbour hosts, 1:
fails; 0 : healthy;}\\ 
1 & 2 & Host Z+ & \\
1 & 3 & Host Y- & \\
1 & 4 & Host Y+ & \\
1 & 5 & Host X- & \\
1 & 6 & Host X+ & \\
\hline
2 & 7-8 & DNP core status & 00: normal; 01: sick; ??\\
\hline
2 & 9-10 & Power Consumption status & 00: normal; 01: warning; 10: alarm\\ 
\hline
2 & 11-12 & Voltage status & 00: normal; 01: warning; 10: alarm\\ 
\hline
2 & 13-14 & Temperature status & 00: normal; 01: warning; 10: alarm\\ 
\hline
2 & 15-16 & Link Z- & \multirow{6}{*}{Link status; 00: normal; 01: sick;
  10: broken}\\
2 & 17-18 & Link Z+ &\\
2 & 19-20 & Link Y- &\\
2 & 21-22 & Link Y+ &\\
2 & 23-24 & Link X- & \\
2 & 25-26 & Link X+ & \\
\hline
5 & 27-31 & & Empty\\
\hline
\hline
\end{tabular}
\caption{DNP local/global watchdog register layout}
\label{tab:DNPwdreg}
\end{table}

\begin{table}[!htb]
\centering
\setlength\extrarowheight{3pt}
\begin{tabular}{|c|c|c|l|}
\hline
\hline
\multicolumn{4}{|c|}{\textbf{Host watchdog register}}\\
\hline 
\hline
\textbf{\# bits} & \textbf{bit range} & \textbf{field name} & \textbf{Description}\\
\hline
1 & 0 & Valid & 1 : register content valid; 0 : register content not valid\\
\hline
2 & 1-2 & Service Net Status & Status of the Service Network; 00:
normal; 01: sick; 10: broken;\\ 
\hline
2 & 3-4 & Memory Status & Status of Memory; 00:
normal; 01: sick; 10: broken;\\
\hline
2 & 5-6 & Peripheral 0 status & Status of Peripheral 0; 00:
normal; 01: sick; 10: broken;\\
\hline
2 & 7-8 & Peripheral 1 status & Status of Peripheral 1; 00:
normal; 01: sick; 10: broken;\\
\hline
23 & 9-31 & & Empty\\
\hline
\hline
\end{tabular}
\caption{Host watchdog register layout}
\label{tab:HOSTwdreg}
\end{table}

\begin{table}[!htb]
\centering
\setlength\extrarowheight{3pt}
\begin{tabular}{|c|c|l|l|}
\hline
\hline
\multicolumn{4}{|c|}{\textbf{Host remote fault descriptor register}}\\
\hline 
\hline
\textbf{\# bits} & \textbf{bit range} & \textbf{field name} & \textbf{Description}\\
\hline
1 & 0 & Z- Service Net Status & Status of the Service Network; 0:
normal; 1: broken;\\ 
1 & 1 & Z- Memory Status & Status of Memory; 0: normal; 1: broken;\\
1 & 2 & Z- Peripheral 0 status & Status of Peripheral 0; 0: normal; 1: broken;\\
1 & 3 & Z- Peripheral 1 status & Status of Peripheral 1; 0:normal; 1: broken;\\
\hline
1 & 4 & Z+ Service Net Status & Status of the Service Network; 0:
normal; 1: broken;\\ 
1 & 5 & Z+ Memory Status & Status of Memory; 0: normal; 1: broken;\\
1 & 6 & Z+ Peripheral 0 status & Status of Peripheral 0; 0: normal; 1: broken;\\
1 & 7 & Z+ Peripheral 1 status & Status of Peripheral 1; 0:normal; 1: broken;\\
\hline
1 & 8 & Y- Service Net Status & Status of the Service Network; 0:
normal; 1: broken;\\ 
1 & 9 & Y- Memory Status & Status of Memory; 0: normal; 1: broken;\\
1 & 10 & Y- Peripheral 0 status & Status of Peripheral 0; 0: normal; 1: broken;\\
1 & 11 & Y- Peripheral 1 status & Status of Peripheral 1; 0:normal; 1: broken;\\
\hline
1 & 12 & Y+ Service Net Status & Status of the Service Network; 0:
normal; 1: broken;\\ 
1 & 13 & Y+ Memory Status & Status of Memory; 0: normal; 1: broken;\\
1 & 14 & Y+ Peripheral 0 status & Status of Peripheral 0; 0: normal; 1: broken;\\
1 & 15 & Y+ Peripheral 1 status & Status of Peripheral 1; 0:normal; 1: broken;\\
\hline
1 & 16 & X- Service Net Status & Status of the Service Network; 0:
normal; 1: broken;\\ 
1 & 17 & X- Memory Status & Status of Memory; 0: normal; 1: broken;\\
1 & 18 & X- Peripheral 0 status & Status of Peripheral 0; 0: normal; 1: broken;\\
1 & 19 & X- Peripheral 1 status & Status of Peripheral 1; 0:normal; 1: broken;\\
\hline
1 & 20 & X+ Service Net Status & Status of the Service Network; 0:
normal; 1: broken;\\ 
1 & 21 & X+ Memory Status & Status of Memory; 0: normal; 1: broken;\\
1 & 22 & X+ Peripheral 0 status & Status of Peripheral 0; 0: normal; 1: broken;\\
1 & 23 & X+ Peripheral 1 status & Status of Peripheral 1; 0:normal; 1: broken;\\
\hline
8 & 24-31 && empty\\
\hline
\hline
\end{tabular}
\caption{Host remote faults descriptor register}
\label{tab:HOSTremfault}
\end{table}

\begin{table}[!htb]
\centering
\setlength\extrarowheight{3pt}
\begin{tabular}{|p{4cm}|p{2cm}|p{2cm}|c|c|p{4cm}|}
\hline
\multirow{2}{*}{\textbf{NAME}} & \multicolumn{2}{c|}{\textbf{DNP-VEP
    reg}} & \multicolumn{2}{c|}{\textbf{\apenetp reg}} & \multirow{2}{*}{\textbf{Description}}\\
\cline{2-5}
&\emph{address}&\emph{mask}&\emph{reg number}&\emph{mask}&\\
\hline
\lofamo DNP local/global watchdog & \texttt{0x00000ec0} &
\texttt{0x07ffffff}&  TBD & TBD & DNP local/global watchdog\\
\hline
\lofamo Host watchdog & \texttt{0x00000ec4} & \texttt{0x000001ff}& TBD
& TBD & Host local watchdog\\
\hline
\lofamo Host remote fault descriptor & \texttt{0x00000ec8} &
\texttt{0x00ffffff} & TBD & TBD & Host remote fault descriptor\\
\hline
Temperature & \texttt{0x00000f40}& \texttt{0x000000ff}& 54 & \texttt{0x0000ff00} & Temperature value\\
\hline
Power consumption & TBD & TBD & TBD & TBD & Power consumption value\\
\hline
Voltage & TBD & TBD & TBD & TBD & Voltage value\\
\hline
Temp thresholds & \texttt{0x000000f00} & \texttt{0xffffffff} & TBD &  \texttt{0xffffffff} & Boundaries for the \emph{normal},
\emph{warning},
\emph{alarm} temperature ranges \\
\hline
Power thresholds & TBD & \texttt{0xffffffff} & TBD &  \texttt{0xffffffff} & Boundaries for the \emph{normal},
\emph{warning},
\emph{alarm} power consumption ranges \\
\hline
Voltage thresholds & TBD & \texttt{0xffffffff} & TBD &  \texttt{0xffffffff} & Boundaries for the \emph{normal},
\emph{warning},
\emph{alarm} temperature ranges \\
\hline
Maxhops & \texttt{0x00000008}& \texttt{0x000000ff}& 46 &
\texttt{0x000000ff} & Set the maximum number of hops\\
\hline
Router exceptions & \texttt{0x00000140}\newline\texttt{0x00000144}&
\texttt{0xffffffff}\newline\texttt{0x00003fff} & 29 & \texttt{0x0000002f} & Router status/exceptions \\
\hline
Channel XP exceptions & \texttt{0x00000440}& \texttt{0x000000ff}& 35 & \texttt{0xfe000000} & Link status/exceptions\\
\hline
Channel XM exceptions & \texttt{0x00000540}& \texttt{0x000000ff}& 36 & \texttt{0xfe000000} & Link status/exceptions\\
\hline
Channel YP exceptions & \texttt{0x00000640}& \texttt{0x000000ff}& 37 & \texttt{0xfe000000} & Link status/exceptions\\
\hline
Channel YM exceptions & \texttt{0x00000740}& \texttt{0x000000ff}& 38 & \texttt{0xfe000000} & Link status/exceptions\\
\hline
Channel ZP exceptions & \texttt{0x00000840}& \texttt{0x000000ff}& 39 & \texttt{0xfe000000} & Link status/exceptions\\
\hline
Channel ZM exceptions & \texttt{0x00000940}& \texttt{0x000000ff}& 40 & \texttt{0xfe000000} & Link status/exceptions\\
\hline
RDMA Exception & \texttt{0x00000240}& \texttt{0x000000ff}& 0 & \texttt{0x00000002}& RDMA status\\
\hline
ENGINE Exceptions & \texttt{0x000000c0} & \texttt{0x00000fff} & N.A. & N.A. & ENGINE status/exceptions\\
\hline
PCI Exceptions & N.A.& N.A.& 9 & \texttt{0xffffffff} & PCI status/exceptions\\
\hline
\end{tabular}
\caption{DNP-VEP and \apenetp registers meaningful for the faults detection.}
\label{tab:FaultDetRegs}
\end{table}
\begin{table}[!htbp]
\centering
\setlength\extrarowheight{3pt}
\begin{tabular}{|l|c|c|c|c|c|c|c|c|c|c|c|c|c|}
\hline
\textbf{Value (hex)} & FF & E4 & D5 & D0 & B2 & 9E & 8A & 80 & 76 & 6C & 62 &
4E & 3A\\
\hline
\textbf{Temp ($^{\circ}$C)} & 127 & 100 & 85 & 80 & 50 & 30 & 10 & 0 & -10 & -20
& -30 & -50 & -70\\
\hline
\end{tabular}
\caption{Temperature values and their encoding for the DNP Temperature
  Register}
\label{tab:TempValues}
\end{table}
\FloatBarrier

\FloatBarrier
\clearpage
\subsection{Fault Detection/Signaling Hypothesis}
\label{FaultDetSignal}
In this section we list the faults and critical events managed by the EURETILE
implementation of the \lofamo design. For each fault listed, we provide two paragraphs: \textbf{1- Fault description:} how it is detected (which component is in
charge of detecting such fault) and \textbf{2- Fault detection:} how the information is conveyed
upwards to obtain the systemic Fault Awareness.\\

\noindent
\begin{longtable}{p{0.2\textwidth}p{0.75\textwidth} }
DNP Link sick & \textbf{Fault description:} A malfunction of the physical
 channel (\eg the cable is damaged, incorrectly plugged in, a source
 of interference is in its range, \etc) can lead to errors during
 packets transmission, resulting in data corruption (commission
 fault).  \apenetp link logic implements Cyclic Redundancy Check with
 the CRC-32 IEEE standard polynomial: for packet sizes from 256 bits up
 to 65792, which are the minimum and maximum sizes that \apenetp
 transmits, this CRC is able to detect a maximum number of
 errors\footnote{The number of errors are here identified by the
 Hamming distance between the corrupted and the uncorrupted messages,
 \ie the number of flipped bits.} that ranges from 6 down to 2\cite{Castagnoli}. The
 error detection is performed by the receiving DNP
 card. A link is considered sick when the ratio between the number of
 errors and the number of packets received by that link overruns a
 given (programmable) threshold.\newline \textbf{Fault detection:} It
 is detected by the DNP links self-test logic on the receiving side,
 that signals the faulty (number of errors over threshold) situation to the DNP fault manager. The
 information is stored in the DNP local/global watchdog register where
 it is caught by the Host during the periodical DNP monitoring
 operated by the local Fault Supervisor. As the DNP 3D network is
 affected by this fault the local FS can communicate the faulty status
 to the other nodes via the Service Network.
\\ \\ 
DNP Link broken &
 \textbf{Fault description:} The physical channel is severed (\eg
 cable unplugged or broken) or there is a total failure of the link logic on
 (just) one the two transferring sides (RX, TX) preventing the message
 to be transmitted (omission fault). This situation can be diagnosed by
 the link logic (when correctly operating) because the operativity of
 the physical channel implies a handshaking protocol between the RX
 and TX sides. As a consequence, a broken cable can be detected by
 both the receiving and transmitting DNP.\newline
 \textbf{Fault detection:} It is detected by the DNP links self-test
 logic both on the receiving and transmitting side, that signal the
 faulty situation to the DNP fault manager. The information is stored
 in the DNP local/global watchdog register where it is caught by the
 Host during the periodical DNP monitoring operated by the local Fault
 Supervisor. As the DNP 3D network is affected by this fault the FS
 can communicate the faulty status to the other nodes via the Service
 Network.
\\ \\ 
DNP Temperature\newline over/under threshold & \textbf{Fault
 description:} \apenetp provides a sensor measuring the FPGA die
 temperature. This information can be both routed to
 the board's control devices and to into the FPGA itself. At the
 moment a control
 logic is implemented as part of the DNP core to convey this
 temperature information to a DNP register encoded as shown in
 table \ref{tab:TempValues}. Four thresholds can be defined for the
 temperature on the basis of the reference temperature limits for
 correct operativity, in order to set the boundaries for the ranges of:
 \emph{normal} operativity, \emph{warning} needed for over/under conditions and \emph{alarm}
 raised. The \lofamo design makes \apenetp able to check the
 temperature and detect and signal an over-under-threshold
 situation.\newline \textbf{Fault detection:} The DNP Local Fault
 Manager checks the DNP temperature status and
 therefore periodically sets (together with the other DNP Status
 information ) the Temperature Status fields in the DNP Local/Global
 register, which is periodically monitored by the Host Fault Manager
 (see Watchdog mechanism specification in section \ref{Watchdog}). In
 this way the Host Fault Manager can become aware of the warning or
 alarming temperature status of the local DNP and send messages to the
 Fault Supervisor through the Service Network (see figure
 \ref{fig:LOFAMO_Temp}).  \\ \\
DNP Power over/under threshold & \textbf{Fault description and Fault Detection:} Similar to the DNP Temperature over/under threshold case.\\ \\
DNP Voltage\newline over/under threshold & \textbf{Fault description and Fault Detection:} Similar to the DNP Temperature over/under threshold case.\\ \\
\end{longtable}
\begin{figure}[!ht]
\centering
\includegraphics[width=0.7\textwidth]{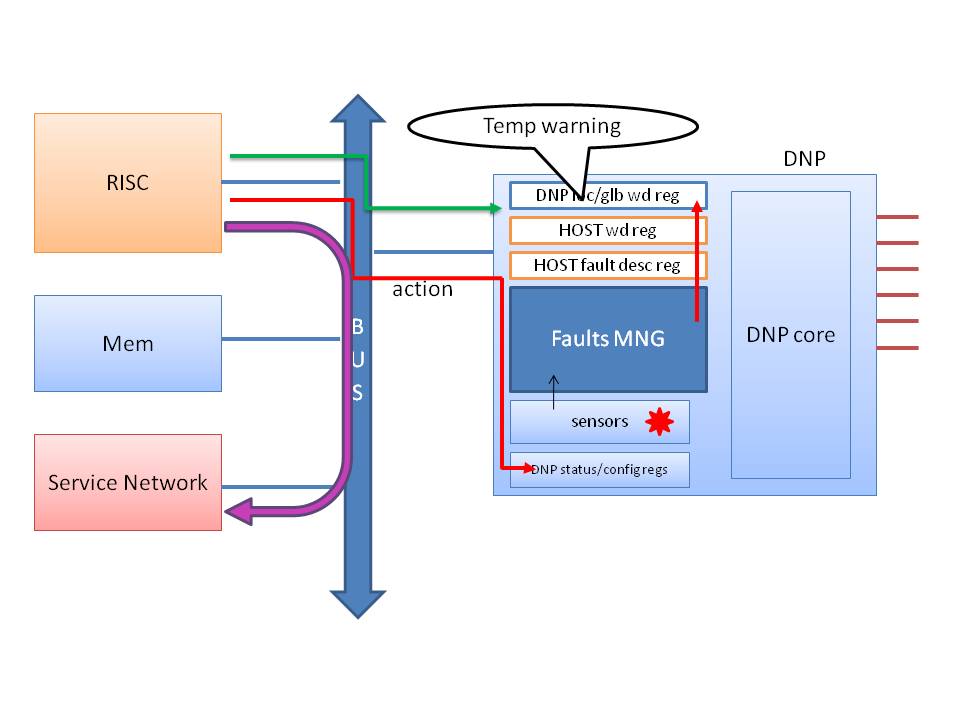}
\caption{Temperature over/under-threshold fault detection and signaling.}
\label{fig:LOFAMO_Temp}
\end{figure}
\FloatBarrier
\begin{longtable}{p{0.2\textwidth}p{0.75\textwidth} }

DNP core sick & 
\textbf{Fault description:} One or more fault in the DNP core internal logic that
cause commission. Typical faults involve the routing
logic (\eg a symptom of this kind of fault can be for example an
\emph{overthreshold number of
hops}) or the RDMA engine. Each of this faults has a related
Exception Register in the DNP register file, so we consider sick a DNP
that has raised an exception.\newline
\textbf{Fault detection:}In
 case that one or more faults in the DNP core logic that cause
 exceptions, the DNP Local Fault Manager set fields in
 the DNP Local/Global Status register. This sick status is
 periodically checked by the Host Fault Manager that can communicate
 this faulty status (and more detailed information that it
 can retrieve by the DNP registers) to other nodes via the Service
 Network.  \\
\\ 
DNP core\newline \emph{meltdown} & \textbf{Fault description:} A fault in the DNP
 core internal logic that causes DNP operativity to be totally and fatally
 compromised.
 This kind of fault can be detected by the DNP fault manager component
 inside the DNP itself in case it is still healthy or by the Host
 thanks to the watchdog
 mechanism described in section \ref{Watchdog}.\newline
\textbf{Fault detection:} A fatal fault
 causes the DNP Local Fault Manager to stop
 its periodic status report in the DNP Local/Global Status
 register. The Host Fault Manager can detect this situation (see Watchdog mechanism specification in section
 \ref{Watchdog}) and signal the fault at global level
 by sending messages via the Service Network (see figure
 \ref{fig:LOFAMO_DNPcore}).\\ \\
\end{longtable}
\begin{figure}[!ht]
\centering
\includegraphics[width=0.7\textwidth]{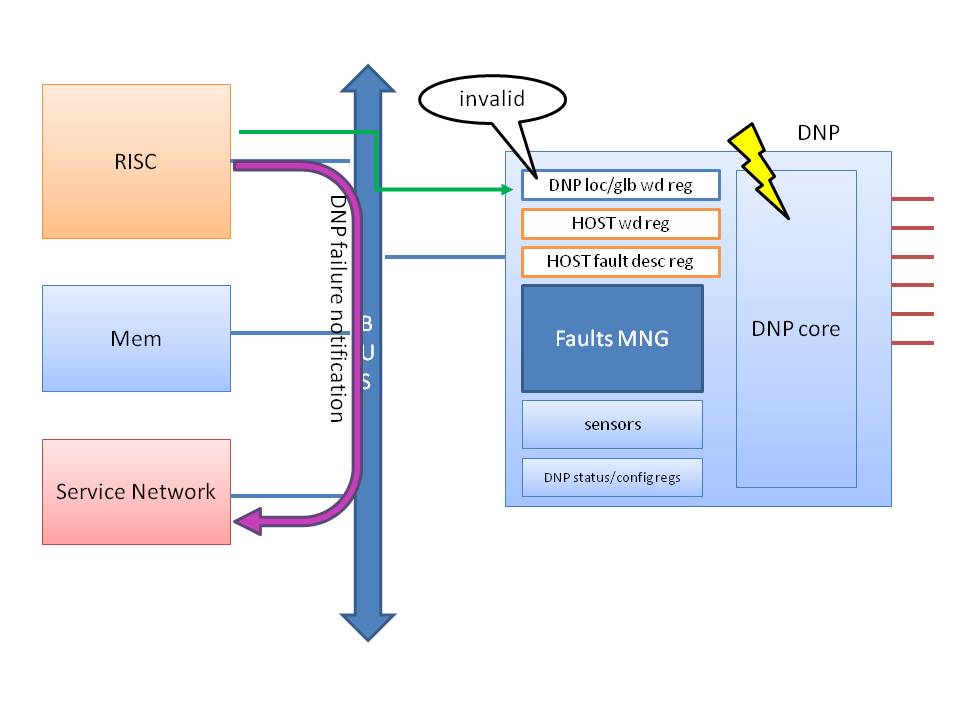}
\caption{DNP core \emph{meltdown} fault detection and signaling.}
\label{fig:LOFAMO_DNPcore}
\end{figure}
\FloatBarrier

\begin{longtable}{p{0.2\textwidth}p{0.75\textwidth} }
HOST Memory, Peripherals (Service Network, \ldots) broken & \textbf{Fault
  description:} Any possible commission fault (on
the Host side) that the Host itself can detect or become aware of. By
definition, these faults are platform-dependent and can be included in
the watchdog fault detection mechanism as described in section
\ref{Watchdog}.\newline
\textbf{Fault detection:} In case one or
more Host Peripherals (DNP excluded) have a fault or are broken the
Host Local Fault Manager should be able to detect the problem and
communicate the faulty status to the upper hierarchy layers of the Fault Supervisor
via the Service
Network, when not faulty, or the 3D network. 
To use the 3D network diagnostic info path, the Host Local Fault
Manager writes the Host Watchdog register signaling the faulty
devices. The DNP Fault Manager that periodically checks this register
becomes aware of the faults and prepare a diagnostic info packet to be
sent through the 3D network. Once received by the neighbouring DNPs
the information is reported as follows: a bit is raised in the Global
fields of the DNP Watchdog register showing the direction of the
faulty neighbour node; the proper field is set in the Host Remote
Fault Descriptor register to convey the type of fault and/or the
device affected by it.\\
\\
Total HOST breakdown and/or bus broken & \textbf{Fault description:}
  Any fault on the Host side that causes a Host omission failure. In
  this category we also include Bus omission faults because from the
  DNP point of view a broken Bus and a completely non operating Host
  do not differ, as both these situations are detected by the watchdog
  mechanism as a lack of activity from the Host side (see section
  \ref{Watchdog}).\newline 
\textbf{Fault detection:} A fault that
  jeopardizes the ability of the Host to update the Host Watchdog
  register is easily detected by the DNP Fault Manager that reads a
  not valid status on that register and sends diagnostic a diagnostic
  packet to the DNP's first neighbours. The information is received
  and processed by the neighbouring DNP Fault Managers marks in their
  DNP watchdog registers the fields corresponding to the faulty
  node. In this way the Host Fault Manager that monitors that register
  can become aware of the situation and send packets to upper
  hierarchy layers of the Fault Supervisor via the Service Network
  (see figure \ref{fig:LOFAMO_host}).\\ \\
\end{longtable}
\begin{figure}[!htb]
\centering
\includegraphics[width=0.8\textwidth]{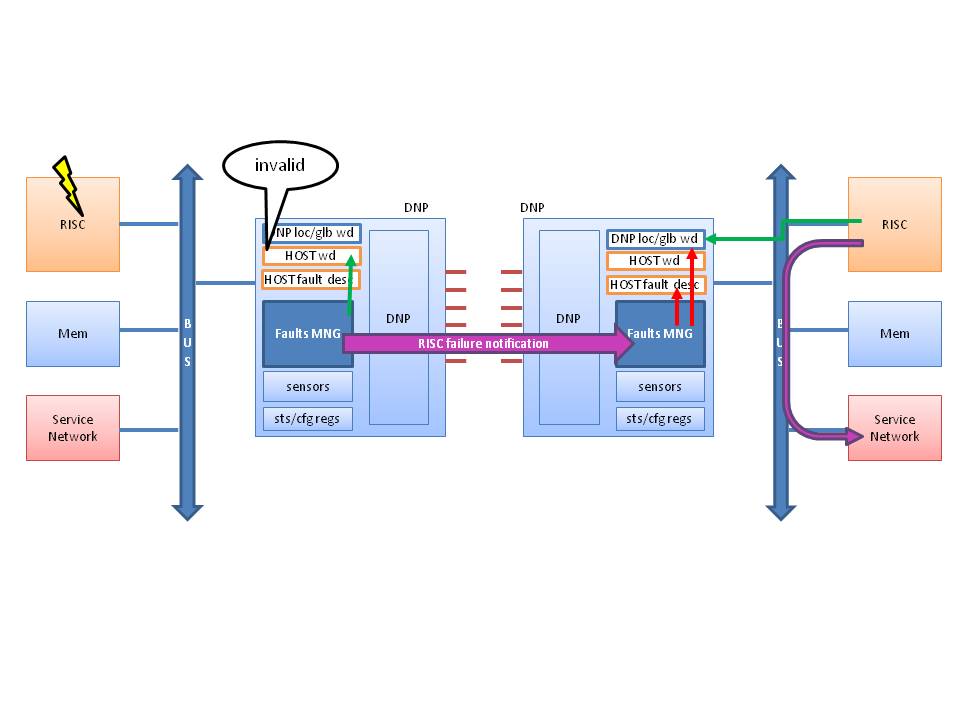}
\caption{Host breakdown detection and signaling.}
\label{fig:LOFAMO_host}
\end{figure}

\FloatBarrier

\section*{Acknowledgment}

This work was supported by the EU Framework Programme 7
project EURETILE under grant number 247846.



%

\end{document}